\documentclass[10pt,conference,letterpaper]{IEEEtran}

\usepackage{graphicx}
\usepackage{amssymb}
\usepackage{amsmath}
\usepackage{algorithm}
\usepackage{algorithmic}
\usepackage{balance}

\newcommand{\mynotex}[1]{}

\IEEEoverridecommandlockouts

\usepackage{tikz}
\usepackage{textcomp}
\usepackage{hyperref}
\usepackage{lipsum}

\newcommand\copyrighttext{%
  \footnotesize \textcopyright 2019 IEEE. Personal use of this material is permitted. Permission from IEEE must be obtained for all other uses, in any current or future media, including reprinting/republishing this material for advertising or promotional purposes, creating new collective works, for resale or redistribution to servers or lists, or reuse of any copyrighted component of this work in other works. Accepted in 2nd Workshop on Cryptocurrencies and Blockchains for Distributed Systems (CryBlock 2019), in conjunction with IEEE INFOCOM, April 29 - May 2, 2019, Paris, France.
}
\newcommand\copyrightnotice{%
\begin{tikzpicture}[remember picture,overlay]
\node[anchor=north,yshift=0pt] at (current page.north) {\fbox{\parbox{\dimexpr\textwidth-\fboxsep-\fboxrule\relax}{\copyrighttext}}};
\end{tikzpicture}%
}

\begin{document}

\title{
Interledger Smart Contracts for Decentralized Authorization to Constrained Things\thanks{This research  has been undertaken in the context of project
SOFIE (Secure Open Federation for Internet Everywhere), which has received funding from EU's Horizon 2020 programme, under grant agreement No. 779984.}
\vspace{-0.0in}
\vspace{-0.1in}
}

\author{
Vasilios A. Siris, Dimitrios Dimopoulos, Nikos Fotiou, Spyros Voulgaris, George C. Polyzos \\
Mobile Multimedia Laboratory, Department of Informatics \\
School of Information Sciences \& Technology \\
Athens University of Economics and Business, Greece\\
\{vsiris, dimopoulosd, fotiou, voulgaris, polyzos\}@aueb.gr
\vspace{-0.1in}
}

\maketitle
\copyrightnotice
\normalfont

\begin{abstract}
We present models that utilize smart contracts and interledger mechanisms to provide decentralized authorization for constrained IoT devices. The models involve different tradeoffs in terms of cost, delay, complexity, and privacy, while  exploiting key advantages of smart contracts and  multiple blockchains that communicate with interledger mechanisms. These include immutably recording hashes of authorization information and policies in smart contracts, resilience through the execution of smart contract code on all blockchain nodes, and cryptographically linking transactions and IoT events recorded on different blockchains using hash and time-lock mechanisms.
The proposed models are evaluated on the public Ethereum testnets Rinkeby and Ropsten, in terms of execution cost (gas), delay, and reduction of data that needs to be sent to the constrained IoT devices.
\end{abstract}


\vspace{-0.01in}
\section{Introduction}
\vspace{-0.01in}
Blockchains have different properties, e.g., Ethereum is a public and permissionless blockchain with  Turing-complete smart contracts and an intrinsic cryptocurrency. Being public, Ethereum, as Bitcoin, provides large-scale decentralized trust at the expense of  high computation costs, hence high transaction fees, and  high transaction delays. On the other hand, permissioned blockchains or distributed ledgers such as Hyperledger Fabric and R3 Corda operate with a restricted set of peers and incur a smaller cost and delay, while supporting different levels of write and read access.
Combining multiple blockchains to implement different functionality, such as authorization and payment, allows different cost, delay, complexity, and privacy tradeoffs. However, multiple blockchains or distributed ledgers must be interconnected in a way that securely binds the transactions on  different ledgers.

Authorization for constrained IoT devices (Things), with limited  connectivity and computation power, requires  support from an \emph{authorization server} (AS)~\cite{Ger++18}; relevant use cases range from authorization for home door locks and health monitoring devices to smart container tracking and industrial actuator control~\cite{Sei++16}.
Offloading authorization functionality from IoT devices to an AS also facilitates the collective management of authorization policies.
However, providing authorization  through a single AS is vulnerable to server failures and misbehavior.
The goal of this paper is to propose and investigate models for providing decentralized authorization with multiple ASes that utilize two blockchains, one for  authorization and the other for payments, in order to reduce the transaction cost and delay compared to a single (public) blockchain.


The proposed models are general and can exploit blockchain and smart contract technology in the  context of authorization in constrained IoT environments. Our realization of the models considers  the OAuth 2.0 delegated authorization framework, which  is based on \emph{access tokens}  and  is a widely used  IETF standard, currently being investigated for authorization in IoT environments by IETF's  Authentication and Authorization for Constrained Environments (ACE) Working Group \cite{Sei++19}. We also consider the CBOR (Concise Binary Object Representation) Web Token (CWT) format~\cite{Sei++19,Jon++18},  a recently proposed standard for compactly encoding \emph{claims} in access tokens which is more efficient than the JSON Web Token (JWT) format.

The contributions of the paper are the following:
\begin{itemize}
\item We investigate interledger mechanisms for securely linking transactions on two   blockchains, to reduce the execution cost and delay compared to  a single (public) chain.
\item We propose an approach for decentralized authorization in constrained IoT environments that utilizes multiple authorization servers (ASes) and  includes  two mechanisms for reducing the amount of data that needs to be sent to the constrained IoT devices (Things).
\item We evaluate the proposed models on two public Ethereum testnets, Rinkeby and Ropsten, in terms of  execution cost (gas),  delay, and reduction of the amount of data that needs to be sent to  IoT devices.
\end{itemize}
Previous work on decentralized authorization based on smart contracts  assumes that IoT devices communicate with the blockchain or  are capable of executing public/private key cryptographic functions.
To the best of our knowledge, this is the first work to investigate the application of smart contracts and multiple blockchains with interledger mechanisms for decentralized authorization to constrained IoT devices.

The remainder of the paper is structured as follows: In Section~\ref{sec:authorization} we present some background on authorization in constrained IoT environments. In Section~\ref{sec:blockchains} we present the proposed models for  decentralized authorization  and in Section~\ref{sec:evaluation} we evaluate the  models.
Finally, in Sections~\ref{sec:related} and~\ref{sec:conclusions} we present related work and  future research, respectively.

\section{Authorization in constrained environments}
\label{sec:authorization}
\vspace{-0.02in}

OAuth 2.0 is a framework for delegated authorization to access a protected resource \cite{Har++12}. It enables a third party application (client) to obtain access with specific permissions to a protected resource, with the consent of the resource owner. Access to the resource is achieved through access tokens, which are
created by an authorization server (AS).
The specific format of the access tokens, which are discussed below, is opaque to the clients and to OAuth 2.0.
The  consent for authorization by the resource owner is provided after the owner is authenticated.
Authorization is provided for different levels of access,  which are termed \emph{scopes}, and for a specific time interval.
The OAuth 2.0 authorization flows can involve intermediate messages exchanged before the  access token is provided by the AS. However, the details of the authorization flow does not impact the general approach of the proposed models, hence in our discussion we only consider the initial client request and the AS's response with the access token.

One type of access tokens are \emph{bearer tokens}. Bearer tokens allow the holder (bearer), independently of its identity, to access the protected resource.
Bearer tokens are the default for OAuth 2.0, which assumes secure communication between the different entities using TLS (Transport Layer Security).
OAuth 2.0 also assumes that the protected resource is always connected to the Internet, hence can  communicate with the AS to check the validity and scope of the access tokens presented by the clients.
Meeting the above two requirements is not always possible in constrained environments \cite{Sei++16}.


JSON Web Token (JWT) is an open standard that defines a  compact format for transmitting claims   as  JSON objects \cite{Jon++15}.
JWTs can use the JSON Web Signature (JWS) structure to  digitally sign or integrity protect claims with a Message Authentication Code (MAC) \cite{Jon++15b}.
Hence, unlike  bearer tokens, JWT/JWS tokens are self-contained, i.e., they include  all the  information for the  resource to verify their integrity without communicating with the AS.
The JWT format is  considered by the W3C Credentials Community Group for \emph{Verifiable Credentials}, which can be combined with \emph{Decentralized Identifiers} or \emph{DIDs}~\cite{Spo++19}.
A more efficient encoding  derived from JWTs but  based on CBOR (Concise Binary Object Representation) is the  CBOR Web Token (CWT)~\cite{Sei++19,Jon++18}, which can be extended to create and process signatures, MACs, and encrypted data~\cite{Sch17}.

In constrained environments, in addition to limited connectivity, the communication between the client and the protected resource is not secure, hence  transmitting bearer tokens or self-contained JWTs/CWTs over such insecure links make them a target for eavesdropping.
To avoid this   Proof-of-Possession (PoP) tokens are used~\cite{Sei++19}. PoP tokens include a normal access token, such as a JWT/CWT, and a PoP key: 
access to the protected resource requires both the  access token and the PoP key, which is  used to secure the link between the client and the IoT device.
The implementation of the decentralized authorization models presented in this paper adopts the CWT format, proposing two  schemes to further reduce the amount of data that needs to be transmitted to the  constrained device.

\vspace{-0.03in}
\section{Blockchain-based authorization}
\label{sec:blockchains}
\vspace{-0.03in}
The advantages from combining authorization based on  frameworks such as OAuth 2.0 with blockchain and smart contracts are the following:
\begin{itemize}
\item Blockchains can immutably record hashes of the information exchanged during authorization and cryptographically link authorization grants to payments and  other IoT events recorded on the blockchain. These records serve as indisputable receipts  in the case of disagreements.
\item Smart contracts can encode authorization policies in an immutable and  transparent manner. Policies can depend on payments as well as  other IoT events that are recorded on the same or on different blockchains.
\item Smart contracts run on all nodes of a blockchain. Hence, sending resource access requests to smart contracts  can protect against DoS attacks that involve a very high resource request rate, since requests are not handled by one node, which would be a single point of failure.
\end{itemize}
We present four models that allow different tradeoffs in terms of  cost, delay, complexity, and privacy:
\begin{itemize}
\item Linking authorization grants to blockchain payments
\item Smart contract handling authorization requests
\item Smart contract and two blockchains for  authorization and payment with interledger mechanisms
\item Decentralized authorization with multiple ASes
\end{itemize}
The first two models correspond to our baseline scenarios: in the first, only hashes of authorization information are immutably recorded on the blockchain and smart contracts are  not used, whereas the second model utilizes  a smart contract but on a single (public) blockchain. The third model, which focuses on our first contribution,  exploits two blockchains whose  transactions are securely linked using interledger mechanisms and quantifies the significant cost reduction that can be achieved by moving smart contract authorization functionality to a permissioned or private  blockchain. The fourth model focuses on both key  contributions of the paper: decentralized authorization for constrained IoT devices utilizing two blockchains with interledger mechanisms.

A hash-lock is a cryptographic lock that can be unlocked by revealing a secret whose hash is equal to the lock's value $h$. Unlocking a hash-lock can be one of the conditions for performing a transaction or for executing a smart contract function.
On a single blockchain, a hash-lock  can be linked to an off-chain capability, e.g.,  message decryption, if the hash-lock secret  is the secret key that can decrypt the message.

Hash-locks can be used on two or more blockchains that support the same hash function, to link a  transaction on one chain to a transaction on the other chain: if the two transactions have hash-locks with the same value, then unlocking one hash-lock would reveal the secret that unlocks the other; hence, the two transactions are cryptographically linked through a dependence relation. More generally, hash-locks combined with AND/OR logic operators can implement elaborate dependencies involving transactions on multiple chains.

Time-locks are blockchain locks  that can be unlocked only after an interval has elapsed. This interval can be measured in absolute time or  in  the number of blocks mined after a specific block.
One usage of time-locks are refunds:  a user (payer) can make a deposit to a smart contract address. The smart contract can have a function, which typically also includes a hash-lock, for a second user to transfer the deposit to another account (the payee's account).
However, if the second user never calls this function, then the first user's deposit could be locked indefinitely in the smart contract's account. To avoid this, the smart contract can also include a refund function that allows the first user to transfer the amount he/she deposited  back to the his/her account; however, this function can be called only after some time interval, which is the interval in which the second user must transfer the deposit from the smart contract account to the payee's account.

Contracts that include both hash and time-locks are referred to as hashed time-lock contracts (HTLCs)~\cite{HTLCs}. HTLCs have been used for atomic cross-chain trading  (atomic swaps) \cite{atomics,But16} and for  off-chain transactions between trustless parties \cite{Poo++16}.
HTLCs can be implemented in blockchains with simple scripting capabilities, such as the Bitcoin blockchain, without  requiring the advanced functionality of smart contracts.
Smart contracts do not increase the capabilities of \emph{interledger} mechanisms based on hash and time-locks, but increase the \emph{intra-ledger} functionality. We investigate these features for decentralized authorization to constrained IoT devices.

In all the models presented below, the client sends a resource access request to the URL of the AS  (model 1) or to the address of the smart contract responsible for handling access to the IoT device (models 2, 3, and 4). The URL or smart contract address can be obtained by  the client sending a query to the IoT device or reading a QR code on it.
However,  this approach  cannot ensure that the legitimate URL or smart contract address is provided by the IoT device. This can be ensured if the client uses a registry service that resides on the blockchain 
and contains a binding between the IoT device's URI and the URL of the AS or the smart contract address handling authorization, or by including this information in  Decentralized Identifier (DID) documents~\cite{Ree++18}.


Finally, in all models we assume that  the client, the resource owner, and the ASes have an account (public/private key pair) on the blockchain (both the authorization and the payment blockchains for models 3 and 4).

\mynotex{
Models:
\begin{itemize}
\item \#1: Link authorization to payment. Immutably record information from OAuth2 message exchange. Payment contract uses time-locks.
\item \#2: Use smart contract to encode authorization policies, prices, etc. Client interacts directly with smart contract. Can still have single blockchain.
\item \#3: Use two blockchains: one is payment blockchain and other is authorization blockchain. Transactions on two blockchains linked using hash time-locked mechanism. Authorization blockchain can be private/permission for improved privacy compared to public/permissionless.

\end{itemize}

}

\vspace{-0.02in}
\subsection{Model \#1: Linking authorization grants to  payments and recording authorization information on the blockchain}
\label{sec:model1}
\begin{figure}[tb]
\centering
\includegraphics[width=3.5in]{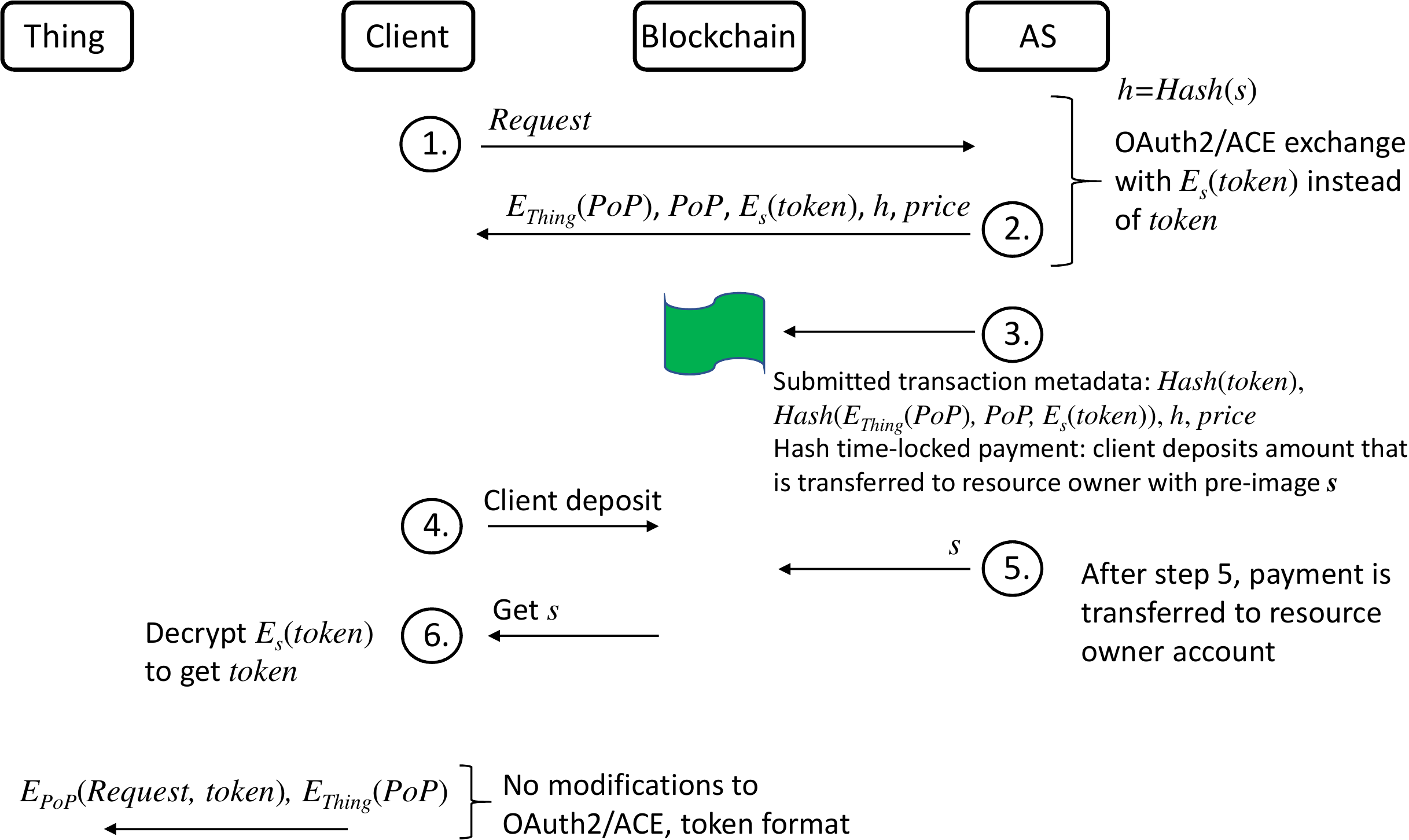}
\vspace{-0.2in}
\caption{Model 1: Authorization grants are linked to blockchain payments and the hashes of the authorization information exchanged are recorded on the blockchain for verification in the case of disputes.}
\label{fig:model1}
\vspace{-0.2in}
\end{figure}

\vspace{-0.02in}
\mynotex{
Key points:
\begin{itemize}
\item Immutable record information exchanged during normal OAuth2 message exchange, that provide verification of authorization grants and non-repudiation in  case of disputes.
\item Either the cleartext of the information exchanges is published or a hash of the information, which represents a tradeoff between transparency and privacy. Level of privacy is also related to whether a public or private/permissioned blockchain is considered.
\item What exactly does recording above information guarantee: 1) client cannot deny receiving specific information contained in OAuth2 message, 2) authorization server cannot deny sending information contained in OAuth2 message.
\item Different access levels can correspond to different prices.
\item Doesn't guarantee that access token client receives provides access to resource.
\end{itemize}
}
With this model the initial communication between the client and the authorization server (AS) follows the normal authorization message exchange, such as  OAuth 2.0, Figure~\ref{fig:model1}.
%
Specifically, in step 1 the client requests resource access from the AS. The AS generates a random  PoP  key which it sends to the client\footnote{The communication link between the client and the AS is secured, hence the PoP key cannot be obtained through eavesdropping.} together with its encryption  with the secret key\footnote{The secret key that the Thing and AS share is established during the provisioning (or commissioning) phase, when the Thing is bound to the AS.} $K_{Thing}$ shared by the Thing (IoT device) and the AS; the client will later use the PoP key to establish a secure communication link with the Thing. Also, the AS sends to the client the access token encrypted with a secret $s$, i.e., $E_s(token)$, the hash $h=Hash(s)$ of the secret $s$, and the price for the requested level of resource access. The secret $s$ is a secret randomly generated by the AS and is required for the client to decrypt $E_s(token)$ and obtain the access token; the AS will reveal the secret $s$ once it confirms that the payment for resource access has been committed on the blockchain.
Communicating the price from the AS to the client allows different  levels of resource access, encoded in the access token's scopes, to correspond to different prices.

In  step 3, two hashes are submitted to the blockchain: the first is the hash of the  token that the AS will reveal to the client once payment has been confirmed.
The second is the  hash of three items:   $E_{K_{Thing}}(PoP)$, the PoP key, and $E_s(token)$; the second hash serves as proof of the information that is communicated using OAuth  between the AS and the client.  
Note that the above authorization exchange does not ensure that the access token  the client obtains from the AS  indeed allows  access to the Thing. 

Also in step 3 a hashed time-lock payment  is initiated on the blockchain, which allows the client to deposit an amount equal to the requested price (step 4). This amount will be transferred to the resource owner's account if the secret $s$ (hash-lock) is submitted to the contract by the AS (step 5) within some time interval. If the time interval is exceeded, then the client can request a refund of the amount it deposited.
Once the secret $s$ is revealed,  the client can get $s$ from the blockchain (step 6) and decrypt $E_s(token)$, and thus obtain the access token.
At this point, the client has all the necessary information to request access from the Thing, using normal OAuth 2.0 with the modifications from the ACE framework.


\vspace{-0.01in}
\subsection{Model \#2: Smart contract handling authorization requests }
\label{sec:model2}
\vspace{-0.01in}
%
In the second model a smart contract is used to transparently record prices and other authorization policies defined by the resource owner, which is also the owner of the smart contract. Examples of such policies include permitting resource access to specific clients, determined by their public/private key pairs on the blockchain, and adding dependence of access authorization on IoT events that are recorded on the blockchain.
%

Whereas in the previous model the client and the AS interacted directly, in this model the interaction is through the smart contract;  this is similar to the next model shown in Figure~\ref{fig:model3}, but using a single blockchain for both authorization and payment.
The smart contract code is executed by all blockchain nodes, providing a secure and reliable execution environment; this provides higher protection against DoS attacks, compared to the model in Section~\ref{sec:model1} where resource access requests are sent directly to the AS.

\vspace{-0.01in}
\subsection{Model \#3: Smart contract and two blockchains with  interledger mechanisms}
\label{sec:model3}
\begin{figure}[tb]
\centering
\includegraphics[width=3.5in]{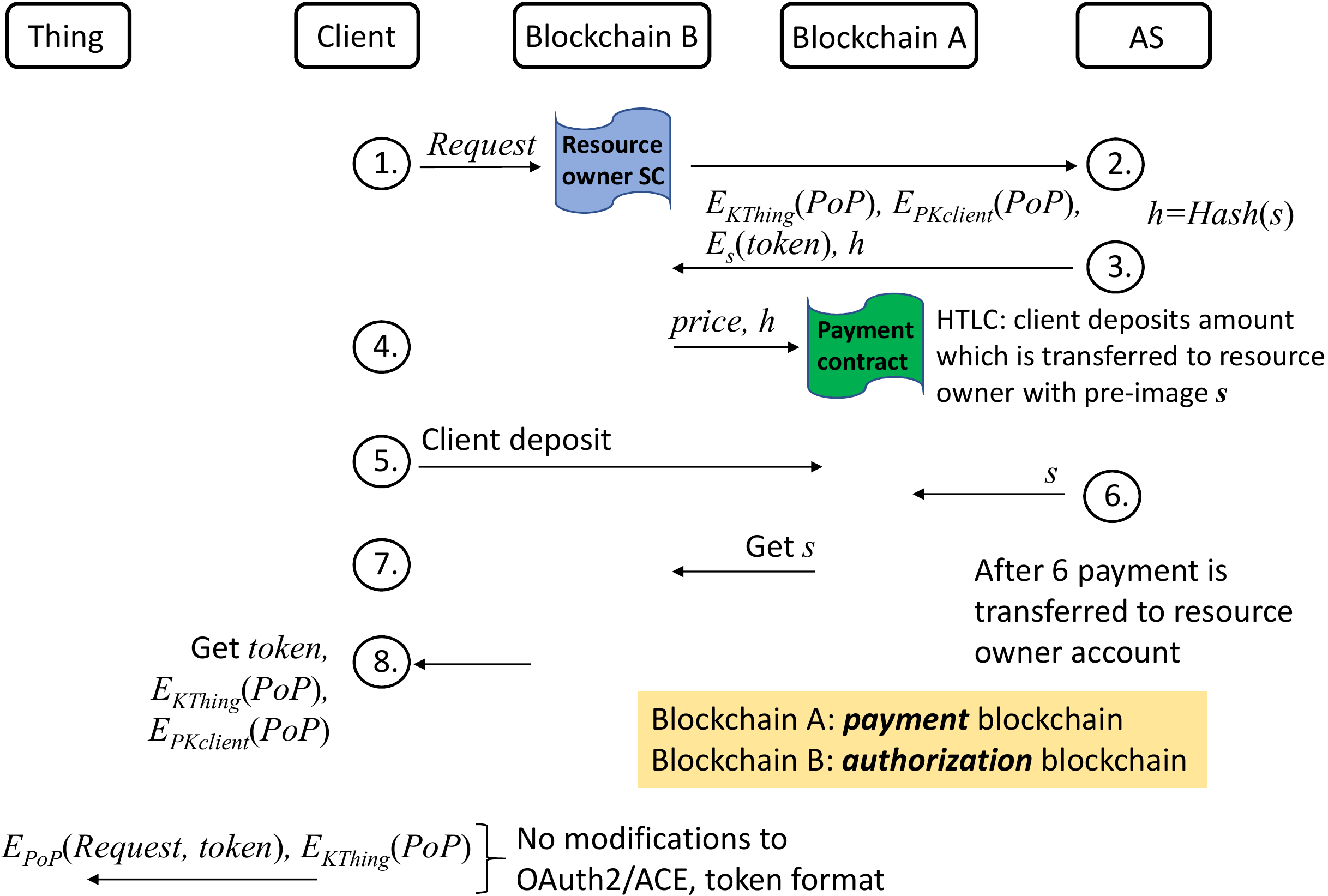}
\vspace{-0.18in}
\caption{Model 3: Smart contract and two chains with interledger mechanisms.}
\label{fig:model3}
\vspace{-0.15in}
\end{figure}

\vspace{-0.01in}
In the model of this subsection the smart contract handling authorization requests and encoding  policies is located on an \emph{authorization blockchain}, while payments for resource access are performed on a \emph{payment blockchain}, see Figure~\ref{fig:model3}.
Depending on whether the authorization chain is a public or a permissioned blockchain, different tradeoffs between transaction cost, delay, and privacy can be realized.

%


A hashed time-lock payment  is initiated on the payment blockchain, where the client can deposit an amount corresponding to the resource access price.
The amount will be transferred to the resource owner's account if the secret $s$ is revealed. Once revealed, the secret $s$ can be submitted to the smart contract on the authorization blockchain, which serves as a record on this blockchain that the payment was successfully performed. The client can obtain the secret $s$ from the authorization blockchain together with the other necessary authorization information to access the protected resource.


One issue with the above model  is how the payment contract on the payment chain is triggered by the resource owner smart contract residing on the authorization chain. One alternative is to have  an \emph{interledger gateway} read the price and hash $h$ from the authorization chain and submit it to the payment chain to initiate a payment (Step 4 in Figure~\ref{fig:model3}), and later read the secret $s$ submitted by the AS on the payment blockchain and record it on the authorization chain (Step 7); the interledger gateway can receive a fee for performing this function. Another alternative is to have this function performed by the AS or  the client.

\vspace{-0.01in}
\subsection{Model \#4: Decentralized authorization with multiple ASes}
\label{sec:model4}
\vspace{-0.01in}
The authorization functionality cannot all be moved onto the blockchain, since it involves processing secret information: keys to produce token signatures  and keys shared with the Thing. Performing  the authorization functions redundantly in the nodes of  a private blockchain would provide a higher level of  resilience to node failures compared to a single AS, but results in reduced security since compromising a single blockchain node would lead to secret keys being disclosed.

Rather than moving all the authorization functionality to the blockchain, we  propose an alternative approach for decentralized authorization that ensures security and provides fault tolerance if some number of ASes are faulty or misbehave.
Let  $n$ be the number of ASes that are collectively responsible for providing authorization. 
Each AS $i$ shares a different  secret key, $K_{Thing_i}$, with the protected resource (Thing).
Authorized access to the Thing requires tokens from $m$ out of $n$ servers. The policy specifying the required number of ASes is defined in the smart contract  and is also known to  the Thing.
Fault tolerance is provided by having $n$ ASes which can respond to requests, but requiring only $m<n$ for authorization to proceed.
Compared to having a single AS, the proposed scheme provides higher security since  $m$ ASes need to agree in order for the client to access the protected resource.

There are two alternatives for how  $m$ servers are selected to provide authorization.
With the first, the smart contract selects the specific $m$ servers; this requires that the smart contract maintains a list of ASes. The list can be updated with information such as the time each AS last responded to an authorization request. Such information allows the smart contract to prioritize ASes in order to select those that respond quickly, hence avoid ASes that have a high delay or are faulty\footnote{Detection of misbehaving ASes that generate incorrect tokens is also possible, but not discussed further due to limited space.}. The evaluation in Section~\ref{sec:evaluation} considers the first alternative.

With the second alternative, the smart contract simply allows all ASes to respond to the authorization request, and selects the first $m$ ASes that respond. With this approach the smart contract does not need to maintain the list of ASes. However,  there is a possibility that the smart contract receives more than $m$ responses. This depends on the duration for mining a block on the blockchain (in the case of public blockchains with Proof-of-Work  consensus) or for obtaining consensus to add it to the blockchain (in the case of permissioned blockchains). In public blockchains, these responses can incur a gas cost independent of whether the ASes that gave the response were among the $m$ ASes to provide decentralized authorization.

\begin{figure}[tb]
\centering
\includegraphics[width=3.5in]{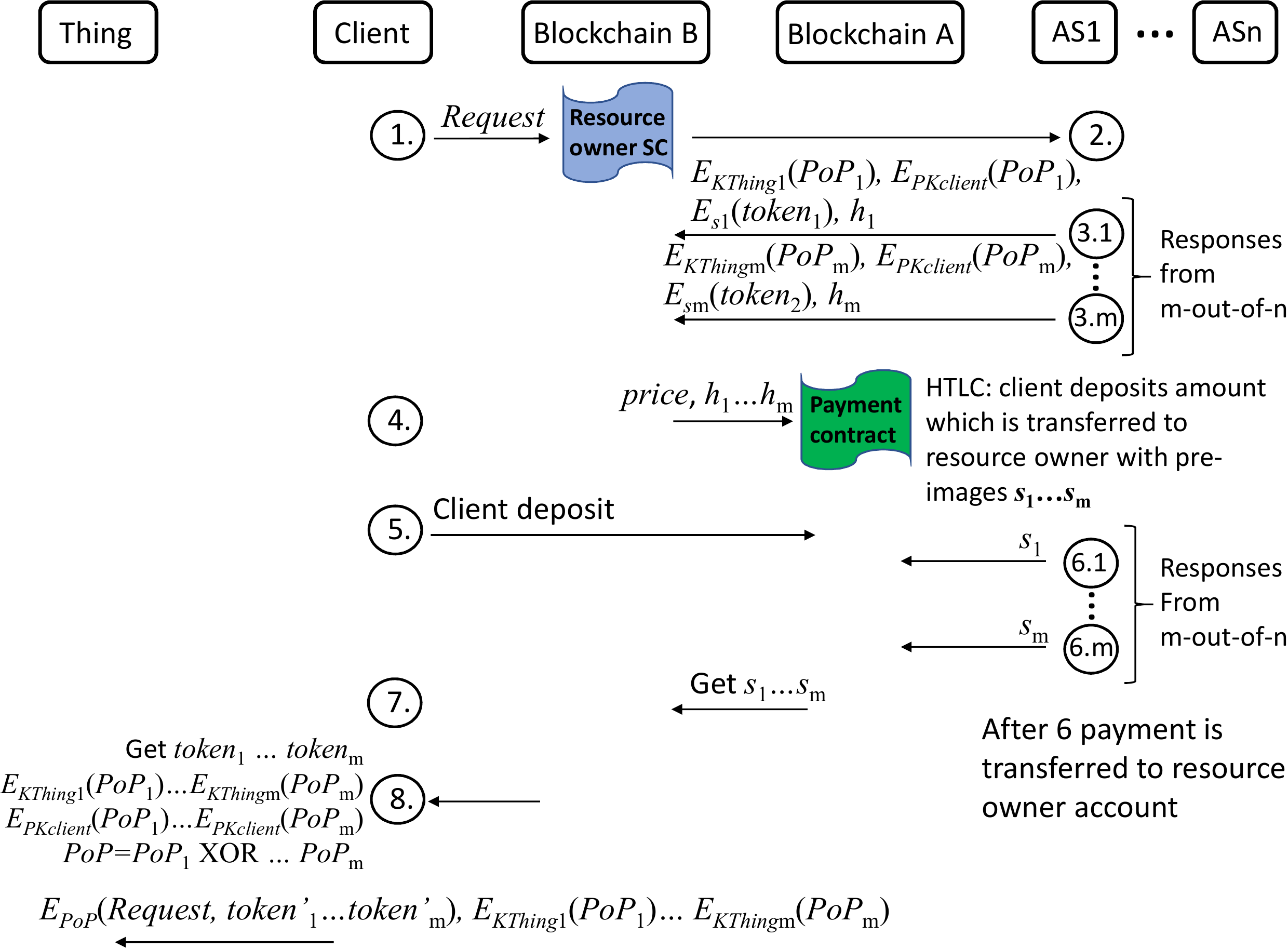}
\vspace{-0.2in}
\caption{Model 4: Decentralized authorization: Each authorization grant requires $m$ out of $n$ AS responses.}
\label{fig:model4}
\vspace{-0.2in}
\end{figure}

In response to the client's authorization request, each AS sends a different PoP key $PoP_i$, encrypted with the Thing's secret key and the client's public key, and an access token with a  MAC tag to ensure its integrity (Figure~\ref{fig:model4}). The client thus obtains $m$ different PoP keys, which it XORs to obtain the secret PoP key that will be used to establish a secure communication link with the Thing. These $m$ PoP keys, encrypted with the Thing's key $K_{Thing_i}$ that it shares  with each of the $m$ ASes, are also sent to the Thing. Hence, if the Thing  performs the same XOR function on the $m$ PoP keys it will obtain the same PoP key as the client.

Recall from the discussion in Section~\ref{sec:authorization} that a requirement is to reduce the amount of data transmitted to  constrained devices. We propose two schemes for reducing the authorization information the client sends to the Thing: aggregate MAC tags and transmission of common token fields once.
With aggregate MAC tags \cite{Kat++08}, the client sends to the Thing the token payloads received from the $m$ ASes, but only one aggregate MAC tag that is computed by taking the XOR of the $m$ MAC tags the client receives from the $m$ ASes.
With the second optimization, the client sends the token fields that are common to all ASes only once (these correspond to  $token'_1, \ldots token'_m$ in Figure~\ref{fig:model3}). The common token fields  include the subject (Thing) the token refers to, the scope of access, the token creation time, the token validity time, and the token type. The fields which are different include the AS and token id fields.
The reduction of the amount of data the client sends to the Thing will be evaluated in the next section.

\vspace{-0.01in}
\section{Evaluation}
\label{sec:evaluation}
\vspace{-0.01in}

For the evaluation we  deployed a local  node  running  Go-Ethereum connected to the Rinkeby public Ethereum testnet\footnote{\texttt{https://www.rinkeby.io/}} and a  node running Parity connected to the Ropsten   testnet\footnote{\texttt{https://ropsten.etherscan.io/}}.
The AS was based on a PHP implementation of the OAuth 2.0 framework\footnote{\texttt{https://github.com/bshaffer/oauth2-server-php}}, extended to support CWT's CBOR encoding\footnote{\texttt{https://github.com/2tvenom/CBOREncode}}. The  client used Web3.js to interact with the blockchain.

In  Table~\ref{tab:results},
the smart contract \& one blockchain and the decentralized \& one blockchain models are equivalent to Figure~\ref{fig:model3} (smart contract \& two blockchains) and  Figure~\ref{fig:model4} (decentralized authorization \& two blockchains) with one blockchain for both authorization and payment. For the results with two blockchains, we use the public blockchain (Rinkeby or Ropsten) as the payment chain and a private Ethereum network as the authorization chain. The results shown  include the gas and delay due to transactions on the public blockchain only.
The results would be similar if we used another technology, e.g., Hyperledger Fabric, as the authorization chain.


\paragraph{Gas}
The second column in Table~\ref{tab:results} shows that the execution cost of a smart contract on the Ethereum Virtual Machine (gas) on the public Rinkeby testnet is significantly higher compared to simply  recording hashes (first line in Table~\ref{tab:results}). However, when the smart contract authorization functionality  is moved to a private blockchain (models with 2 BCs in Table~\ref{tab:results}), then the execution cost is significantly reduced: For 1, 2, and 3 ASes  the execution cost when two blockchains are used is 33.2\%, 23.1\%, and 21.1\% of the execution cost when a single blockchain is used.

\begin{table}[tb]
    \caption{Execution cost (Gas) and delay - Rinkeby} 
    \centering 
\vspace{-0.1in}
\scriptsize{
\begin{tabular}{|c|r|c|} %
        \hline \scriptsize{Model}
        &  \scriptsize{Gas} & \scriptsize{Delay in secs (s)}   \\
        &   & \scriptsize{(95\% conf. int.)}   \\
        \hline 
        \hline
\scriptsize{Hashes of auth. inform.-Fig.~\ref{fig:model1}} & \scriptsize{102\,489} & \scriptsize{43,2  (42,3, 44.1)} \\ \hline 
\scriptsize{SC \& 1 BC} & 258\,166 & 59.3 (57.6, 61.1) \\ \hline 
\scriptsize{SC \& 2 BCs-Fig.~\ref{fig:model3}} & 85\,682 & 43.0 (39.8, 46.2)\\ \hline
\scriptsize{Dec-Auth 2-of-4 \& 1 BC} & 1\,440\,540 & 60.5 (54.4, 67.3) \\ \hline
\scriptsize{Dec-Auth 2-of-4 \& 2 BCs-Fig.~\ref{fig:model4}} & 332\,569 & 42.1 (39.4, 44.8)\\ \hline 
\scriptsize{Dec-Auth 3-of-4 \& 1 BC} & 2\,124\,249 & 63.7 (57.2, 70.2) \\ \hline
\scriptsize{Dec-Auth 3-of-4 \& 2 BCs-Fig.~\ref{fig:model4}} & 447\,940 & 44.7 (39.6, 49.9)\\ \hline
                 \end{tabular} }
    \label{tab:results}
\vspace{-0.2in}
\end{table}

\paragraph{Delay}  The delay is due mainly to the block mining time. The smart contract model with one blockchain has four transactions, while the model that records  only hashes  has three; hence, the    delay for the smart contract model is expected to be 33\% higher; this agrees with the results  in the third column of Table~\ref{tab:results}, according to which  the smart contract model with one blockchain has average delay 59.3s, which is 37.3\% higher than the delay when only hashes are recorded, 43.2s (also shown is the confidence interval from 20 runs).
Table~\ref{tab:results} quantifies the reduced delay  when a public chain is combined with a private chain: e.g., the 2 out of 4 decentralized model with two chains has average delay 42.1s, which is 30.4\% smaller than the delay with one chain, 60.5s.
Table~\ref{tab:results} also shows that for  both  one and two blockchains, the average delay is not significantly influenced by the number of ASes. Also, for two blockchains the average delay is close to the delay when only hashes are recorded.

Table~\ref{tab:results_ropsten} shows that the  delays and  confidence intervals for the Ropsten  testnet are higher than the Rinkeby testnet. We attribute this difference to the fact that Rinkeby uses the Proof-of-Authority (PoA) for distributed consensus, while  Ropsten  uses  Proof-of-Work, as  the  Ethereum mainnet.
For both the Rinkeby and Ropsten testnets, the delay depends on the gas price that is given when a transaction is submitted.

\begin{table}[tb]
    \caption{Delay - Ropsten} 
    \centering 
\vspace{-0.1in}
\scriptsize{
\begin{tabular}{|c|c|} %
        \hline \scriptsize{Model}
        &   \scriptsize{Delay in secs (s)} \scriptsize{(95\% conf. int.)}  \\
        \hline 
        \hline
\scriptsize{Hashes of auth. inform.-Fig.~\ref{fig:model1}} & \scriptsize{53.2  (40.3, 66.1)} \\ \hline 
\scriptsize{SC \& 1 BC}  & 64.4 (52.3, 77.1) \\ \hline 
\scriptsize{SC \& 2 BCs-Fig.~\ref{fig:model3}}  & 57.8 (46.0, 69.7) \\ \hline
\scriptsize{Dec-Auth 2-of-4 \& 1 BC}  & 76.7 (61.5, 92.0) \\ \hline
\scriptsize{Dec-Auth 2-of-4 \& 2 BCs-Fig.~\ref{fig:model4}} & 49.1 (37.7, 60.5) \\ \hline
\scriptsize{Dec-Auth 3-of-4 \& 1 BC} & 77.5 (60.5, 94.6) \\ \hline
\scriptsize{Dec-Auth 3-of-4 \& 2 BCs-Fig.~\ref{fig:model4}}  & 52.2 (42.6, 61.7) \\ \hline
                 \end{tabular} }
    \label{tab:results_ropsten}
\vspace{-0.22in}
\vspace{-0.05in}
\end{table}

\paragraph{Reduction of data client sends to Thing}
Utilizing CWT encoding instead of JWT  reduces the size of  tokens from 310 to 122 bytes. For the smart contract with one blockchain model, the transaction cost is higher by approximately 17\% compared to the cost shown in Table~\ref{tab:results}.

The proposed optimizations further reduce the amount of data that the client needs to send to the Thing. For decentralized authorization with three ASes and without the optimizations, the client  sends 366 bytes ($3 \times 122$ bytes) for three tokens and 96 bytes ($3 \times 32$ bytes) for three PoP keys, a total of 462 bytes. With  aggregate MACs, the client sends one aggregate MAC instead of three, i.e. 64 bytes less, hence a total of 398 bytes, which is a 13.9\% reduction. The optimization where common token fields are sent once results in  84 bytes less, which is a 18.2\% reduction, reducing the size to 314 bytes. The two optimizations together give a 32.0\% reduction of the number of bytes the client needs to send to the Thing. The reduction for more ASes would be higher.

\vspace{-0.012in}
\section{Related work}
\label{sec:related}
\vspace{-0.012in}

The work in \cite{And++17} presents a blockchain-based decentralized authorization system where authorization proofs  can be efficiently verified. The work in \cite{Xu++18} presents a decentralized access control system where IoT devices are required to interact directly with the blockchain and are assumed to be always connected, while
\cite{Mae++17b,Zhang++18} present solutions  where policies and access control decisions are directly recorded on Bitcoin's blockchain.
%
%
%
The work in \cite{Alp++18b2} present a system based on OAuth 2.0 where a smart contract generates authorization tokens, which a key server verifies in order to provide  private keys that allow clients to access a protected resource.
The work in
\cite{Har18} contains a high level description of using smart contracts  with OAuth 2.0 to provide an  architecture where a user can freely select the server to provide authorization for the user's protected resource.
Finally, threshold signatures can be used to achieve authorization from a subset of parties that possess a share of a private signing key~\cite{Bol2002}.

All the above works assume that the IoT devices interact directly with the blockchain or are capable devices, i.e. they are always connected to the Internet and are capable of implementing public/private key cryptographic functions. We do not make these assumptions, and propose a scheme where the authorization function is distributed across multiple servers.
Our previous work \cite{Fot++18} considered the case of a single AS and a single chain, and proposed an approach  to  verify that the  IoT device and AS share a common secret.


\vspace{-0.013in}
\section{Conclusions and future work}
\vspace{-0.013in}
\label{sec:conclusions}

Smart contracts can increase the functionality of a blockchain, and transparently encode authorization policies and logic.
However, smart contracts are costly if  executed on a public blockchain.
Interledger mechanisms enable the interconnection of multiple blockchains, hence allow moving smart contract functionality to private or permissioned blockchains with a lower execution cost.
This paper has proposed  such models for  decentralized authorization to constrained IoT devices and quantified their performance in terms of  reduced transactions costs and delay, while also proposing   mechanisms to reduce the amount of data that needs to be sent to  IoT devices.
We are currently investigating solutions for providing more general services, such as tracking of assets in a supply chain,  in a decentralized  manner over multiple ledgers.

\mynotex{
Smart contracts can increase the functionality that is implemented in a blockchain, and transparently encode authorization policies and logic.
However, smart contracts are costly if they are executed on a public blockchain.
Interledger mechanisms can enable using multiple blockchains, hence moving smart contracts to blockchains with a lower execution costs.

}

\mynotex{
Extensions and related/possible future work includes the following:
\begin{itemize}
\item The decentralized authorization approach presented in this paper has in some cases similarities with the problem of decentralized oracles, which allow smart contracts to interact with the real world, e.g. by accessing off-chain services through APIs. Specifically, if the interaction involves deterministic queries, such as ``Which city is the capital of Greece?'', then the problem is similar. If however the interaction involves non-deterministic queries, such as ``What is the current temperature in Athens?'', the problem is different.
\item Off-chain transactions to reduce the cost of multiple intermediate payments.
\item Delegated authorization supporting out-of-order authorization,  authorization grants with limited duration; tradeoffs between limited duration authorization grants and revocation; Work on delegated authorization and blockchains is contained in \cite{And++17}.
\end{itemize}
}



\balance
\bibliographystyle{IEEEtran}

{
\bibliography{IEEEabrv,auth} }

\end{document}